\documentclass[aip]{revtex4-1}
\usepackage{graphicx}

\begin{document}

% Use the \preprint command to place your local institutional report number
% on the title page in preprint mode.
% Multiple \preprint commands are allowed.
%\preprint{}

\title{Combined neutron reflectometry and rheology} %Title of paper

\author{M.~Wolff}

\email[]{max.wolff@physics.uu.se}

%\homepage[]{Your web page}

%\thanks{}

%\altaffiliation{}

\author{P.~Kuhns}

\affiliation{Division for Materials Physics, Department of Physics and Astronomy, Uppsala University, Box 516, 751 20, Uppsala,
Sweden}

\author{G.~Liesche}

\affiliation{Hochschule Bremerhaven}
\affiliation{Institut Laue Langevin, BP 156, 38042, Grenoble, France}

\author{J.~F.~Ankner}

\affiliation{Spallation Neutron Source, Oak Ridge National Laboratory, Oak Ridge, Tennessee, USA}

\author{J.~F.~Browning}

\affiliation{Spallation Neutron Source, Oak Ridge National Laboratory, Oak Ridge, Tennessee, USA}

\author{P.~Gutfreund}

\email[]{gutfreund@ill.eu}

\affiliation{Institut Laue Langevin, BP 156, 38042, Grenoble, France}

% Collaboration name, if desired (requires use of superscriptaddress option in \documentclass).

% \noaffiliation is required (may also be used with the \author command).

%\collaboration{}

%\noaffiliation

\date{\today}

\begin{abstract}
We have combined neutron reflectometry with rheology in order to investigate the solid boundary of liquids and polymers under shear deformation.
Our approach allows one to apply a controlled stress to a material while resolving the structural arrangements on the sub nanometer length scale with neutron reflectivity, off-specular and small angle scattering at the same time.
The specularly reflected neutron intensity of a 20 \% by weight solution of the Pluronic F127 in deuterated water in contact to a OTS covered and a Piranha treated silicon wafer is evaluated.
We find a pronounced difference in the structure formed by the polymer micelles at the two surfaces, which is explained by the difference in the affinity of the micellar shell to the solid interfaces.
Under deformation the near interface structure changes at deformations of about 2, 30 and 900 \%.
The structural changes are correlated with changes in the storage and loss modulus of the polymer solution revealing a transition from more solid to more liquid like properties.

\end{abstract}

% insert suggested PACS numbers in braces on next line

\maketitle %\maketitle must follow title, authors, abstract and \pacs

\section{Introduction}

Relating macroscopic properties, like e.g. viscosity, to structure in liquids and viscoelastic materials is an important issue for the understanding and tailoring of viscoelastic materials.
It is known that on a microscopic scale the local shear gradient may differ from the mean value probed in macroscopic experiments \cite{Olmsted:2008p283,Sandrin:2001p2896}, an important issue for shear induced ordering.
These variations in the shear gradient might be present in the volume of the material and then manifest in shear bands or exist close to the surface, resulting in apparent surface slip or stick \cite{Neto:2005p2513,Persson:1998,Bushan:1995p607,Krim:1996p48,Thompson:1997p2933,BARRAT:1999p3156,Campbell:1996p520}.
In any case, the solid-liquid interface represents a singularity to a liquid system and it is known that for static liquids ordering critically depends on the properties of the solid-liquid interface as well as on its topology \cite{Jones:1989p280,Fredrickson:1987p2535,Anastasiadis:1989p1852,Kellogg:1996p2503,Huang:1998p7641,Wolff:2004p2390}.\\
In the past mainly small angle neutron or x-ray scattering (SANS or SAXS) has been used to probe the structure of liquids under shear in a Cuette type of shear cell \cite{Zipfel:1998p683,Escalante:2000p3062,Eiser:2000p2491}.
Spatial sensitivity resulted from a reduction of the beam size determining the resolution of the experiment.
No particular interest was paid to the region of the solid-liquid interface except for the works in which a neutron confinement cell is presented \cite{Kuhl:2001p1715, Smith:2006p2411}.
In contrast to this approach, where a liquid is confined in a narrow gap, our setup combines a standard rheometer with neutron reflectometry.\\
Neutron reflectometry is a well established tool to investigate surface anomalies.
Neutrons are particularly valuable for the study of solid-liquid interfaces, since they are highly penetrating for many engineering materials, allowing the investigation of buried interfaces.
In addition, they are sensitive to light elements, like hydrogen.
The fact that neutrons are scattered from the nuclei of atoms makes them sensitive to different isotopes, allowing contrast variation measurements, by replacing hydrogen with deuterium, for example.
In this context one striking example where neutrons contributed heavily are density profiles of liquids close to the solid surface and more specific the discussion on whether a density depleted water layer may exist close to hydrophobic surfaces.
It was shown that a density depleted layer may, if it exists at all, extend at maximum to a fraction of a nm at certain surfaces \cite{Steitz:2003p3385,Maccarini:2007p3669,Doshi:2005p3307,Seo:2006p3099,Mezger:2006p3205}.
Such a depleted layer will influence the flow properties in the near surface region \cite{VINOGRADOVA:1995p2925}.\\
In order to get additional information on in-plane correlations in the liquid, ordering can be investigated by grazing incident small angle neutron scattering (GISANS) \cite{Buschbaum:2009:107,Buschbaum:2004p207,Wolff:2005p2521}.
Combining diffraction techniques with additional surface sensitivity by use of the grazing incident beam geometry can give direct information to help us relate viscoelasticity to structure in the region very near to the surface (10-100 nm).
For example, for a dilute surfactant solution, chosen to be susceptible to shear ordering, a well ordered hexagonal structure has been observed close to a solid interface under flow \cite{HAMILTON:1994p3305}.\\
When probing polymer melts or complex liquids with oscillatory shear the macroscopic properties, e. g. storage and loss modulus, critically depend on the ratio between the internal relaxation times and length scales as compared to the external strain field applied.
In many materials this interaction results in transitions from more solid to more liquid behavior in different frequency/amplitude regimes.\\
In this article we describe rheology done in situ on neutron reflectometers.
We present data taken with a micellar solution undergoing structural changes toward increasing deformation.
The changes found in the structure occur at specific points in the storage and loss modulus plotted vs. deformation.

\section{Experimental details}

In order to measure the reflectivity of the solid/liquid interface while shearing viscoelastic materials in-situ, a rheometer, \textit{Anton-Paar MCR-501}\footnote{\label{comprod} Commercial materials, instruments and equipment are identified in this paper in order to specify the experimental procedure as completely as possible.
In no case does such identification imply a recommendation or endorsement nor does it imply that the materials, instruments, or equipment identified are necessarily best available for the purpose.}, with a cone/plate geometry was modified to meet the special geometrical requirements of neutron reflectometers.

\subsection{Scattering geometry}

\begin{figure}
\begin{center}
\includegraphics[width=0.5\textwidth]{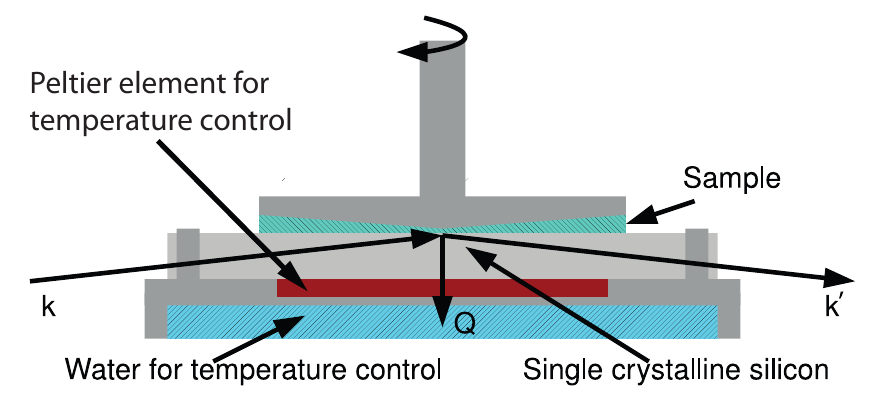}
\caption{\label{scatgeo} Sketch of the experimental set-up. Neutrons enter a single crystalline block of silicon on the narrow side and get scattered from the bottom of the solid-liquid interface \cite{Wolff:2008p11331}.}
\end{center}
\end{figure}

We used a scattering geometry in which the neutron beam passes through the silicon substrate and is reflected downwards from the horizontal solid/liquid interface.
The vertically collimated neutron beam enters ( for dimensions see table \ref{dimensions}) a single crystalline silicon block on the narrow side and reflects from below the liquid sample.
The sample geometry is sketched in Figure \ref{scatgeo} and was described in more detail previously \cite{Wolff:2008p11331, Wolff:2005p2521}.
In the earlier studies shear devices were used, allowing us to apply a constant shear rate.
This made it possible to study changes in the shear viscosity and relate those to changes in the interfacial structure in the liquid.
However, information on the elastic and loss moduli could not be obtained.
In order to close this gap, commercial rheometers (e.g. \textit{Anton-Paar MCR-501})\footnote{\label{comprod} Commercial materials, instruments and equipment are identified in this paper in order to specify the experimental procedure as completely as possible.
In no case does such identification imply a recommendation or endorsement nor does it imply that the materials, instruments, or equipment identified are necessarily best available for the purpose.} have been adapted to meet the specific geometrical requirements of neutron reflectometers.
Two dedicated instruments have been mounted on the sample stages of the Liquids reflectometer at the Oak Ridge National Laboratory (Oak Ridge, TN, USA) and the instrument FIGARO at the Institute Laue-Langevin (Grenoble, France).
\begin{figure}
\begin{center}
\includegraphics[width=0.5\textwidth]{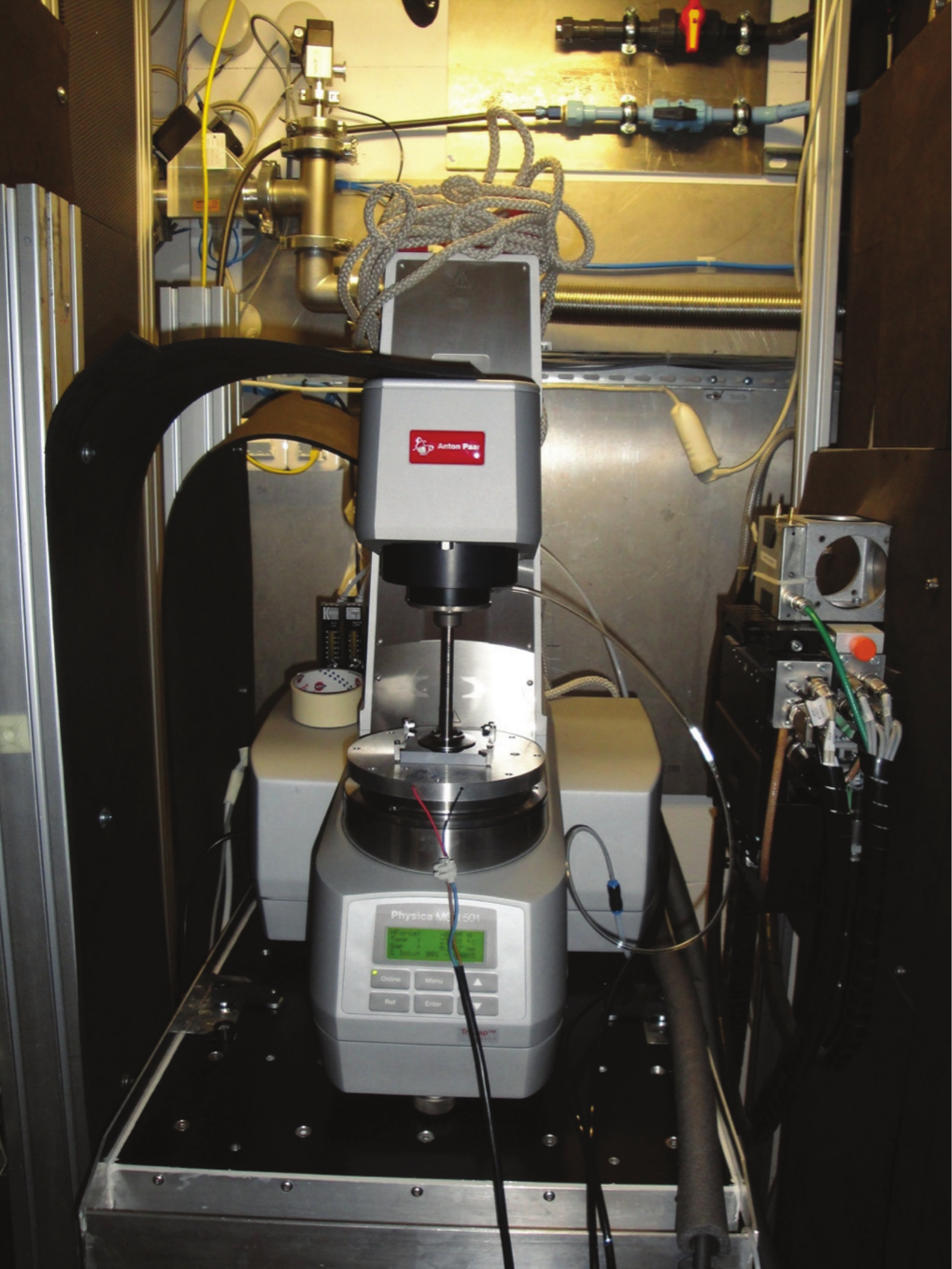}
\caption{\label{photo} Photograph of the experimental set-up. A rheometer \textit{Anton-Paar MCR-501}$^{\ref{comprod}}$ mounted on the reflectometer FIGARO at the ILL, Grenoble, France.}
\end{center}
\end{figure}
The parameters of the two set ups are summarized in table \ref{dimensions}:
\begin{table}
\begin{center}
\begin{tabular}{c|c|c}
						&Figaro			&Liquids reflectometer		\\ \hline \hline
Substrate					&70*70*10 mm$^3$	&2 inch diameter, 10 mm thick     \\ \hline \hline
Rheometer:				&&\\ \hline \hline
Temperature control			&Peltier			&Air flow\\
Temperature range			&				&\\
Maximum torque [mNm]		&300			&230\\
Min. torque, rotation	[nNm]	&10				&10\\
Min. torque, oscillation [nNm]	&2				&2\\
Torque resolution [nNm]		&0.1				&0.1\\
Angular deflection [mrad]		&1 to $\inf$		&1 to $\inf$\\
Angular resolution [nrad]		&10				&10\\
Min. angular velocity	 [rad/s]	&10$^{-9}$		&10$^{-9}$\\
Max. angular velocity [rad/s	&314			&314\\
Max. speed [1/min]			&3000			&3000\\
Min. angular frequency [rad/s]	&10$^{-7}$		&10$^{-7}$\\
Max. angular frequency [rad/s]	&628			&628\\
Normal force range [N]		&0.005-50		&0.005-50\\
Normal force resolution [mN]	&0.5				&0.5\\ \hline \hline
Reflectometer				&				&\\\hline \hline
Wavelength band			&1.7 - 27 \AA				&2.5 - 17.5 \AA\\
Bandwidth				&						&3.5 \AA\\
Q-range					&0.005 - 0.5 \AA$^{-1}$ 		&0 - 0.3 \AA$^{-1}$\\
Typical flux at sample		&10$^{-8}$ ns$^{-1}$cm$^{-2}$	&\\
Angular divergence			&0.005 - 0.1 degree				&\\
Detector					&2 D				&2 D\\
Data format				&Integrated with preselceted time slices		& Time stamped\\
\end{tabular} 
\end{center}
\caption{\label{dimensions} Parameters for the instrumental set up at ILL and SNS.}
\end{table}
Both instruments are equipped with a two dimensional position sensitive detector and allow, in addition to the specularly reflected intensity, the collection of the diffuse scattering as well as grazing incident small angle scattering.
Utilizing these different scattering geometries, it is possible to resolve the near surface structure parallel to the solid-liquid interface on length scales ranging from nm up to 100 $\mu$m \cite{Wolff:2005p2521}.

\subsection{Sample preparation}

The sample, Pluronic F127 (EO$_{99}$-PO$_{65}$-EO$_{99}$), was obtained from BASF Wyandotte Corp. (New Jersey, USA) and used without further purification.
The bulk properties are known in great detail \cite{Wanka:1994p45}.
Pluronics are linear polymers featuring a core of propyleneoxide (PO) and two terminating ethyleneoxide (EO) branches and can self assemble in aqueous solutions \cite{Mortensen:1992p3097}. 
The resulting agreggations have a hydrophobic core and a hydrophilic shell.
At a critical micelle volume fraction the micelles crystallize.
This shear induced crystallization has been observed experimentally by small angle scattering \cite{Mortensen:1996p2979}.
Close to a hydrophilic interface the micelles are adsorbed in layers from the liquid phase as the crystalization transition is approached \cite{Gerstenberg:1998p2465}.
For our experiment, F127 was diluted to 20 \% (by weight) in deuterated water and poured into the sample cell in the liquid phase to prevent shearing of the percolated phase before the measurement.\\
Two functionalized single crystalline silicon (100) substrates (70~mm * 70~mm * 10~mm, optically polished, obtained from \textit{CrysTec}$^{\ref{comprod}}$, Germany) were used for the experiments.
In order to provide high surface energy, one of the two wafers was chemically cleaned in freshly prepared Piranha solution [(50/50 v/v H$_{2}$SO$_{4}$(concentrated) and H$_{2}$O$_{2}$(30~\% aqueous)\footnote{\label{corrosive} Corrosive. Acid-resistant gloves, protective goggles, and lab coats must be worn handling the Piranha solution.}] resulting in a completely wetted surface with interfacial energy of approx. $\gamma_{sl}=70\frac{mJ}{m^{2}}$ \cite{Maccarini:2007p3669}.
The second silicon wafer was cleaned by the same method and then had chemically grafted onto it an octadecyl trichlorosilane (OTS) monolayer, resulting in an interfacial energy of $\gamma_{sl}=19\frac{mJ}{m^{2}}$.\\
Neutron reflectivity measurements with in-situ rheology have been performed on the Liquids reflectometer at the Oak Ridge National Laboratory (Oak Ridge, TN, USA) and the instrument FIGARO \cite{campbell} at the Institute Laue-Langevin (Grenoble, France).
Both are time of flight instruments allowing one to probe a relatively large q-range for one incident angle at the same time, which is advantageous since part of the reflectivity and most of the Bragg scattering resulting from the crystalline order in the polymer solution can be collected without changing the instrument settings.
The data shown below in this manuscript were taken on FIGARO for two incident beam angles of 0.8$^\circ$ and 1.4$^\circ$, taking data for 10 s and 60 s, respectively. The chopper speed was set to 1135 rpm with a phasing of 12.5$^\circ$, 21$^\circ$ and 45$^\circ$ for the second, third and fourth chopper. This settings result in a wavelength resolution of 7.0 \% (FWHM of the Gaussian equivalent). The beam defining slits were set to a width and height of 40 mm and 0.338 mm for the source slit and 25 mm and 0.108 mm for the sample slit. A detailed description of the instrument FIGARO as well as the standard data reduction can be found in ref. \cite{campbell}.\\

\section{Results}

\begin{figure}
\includegraphics[width=450pt]{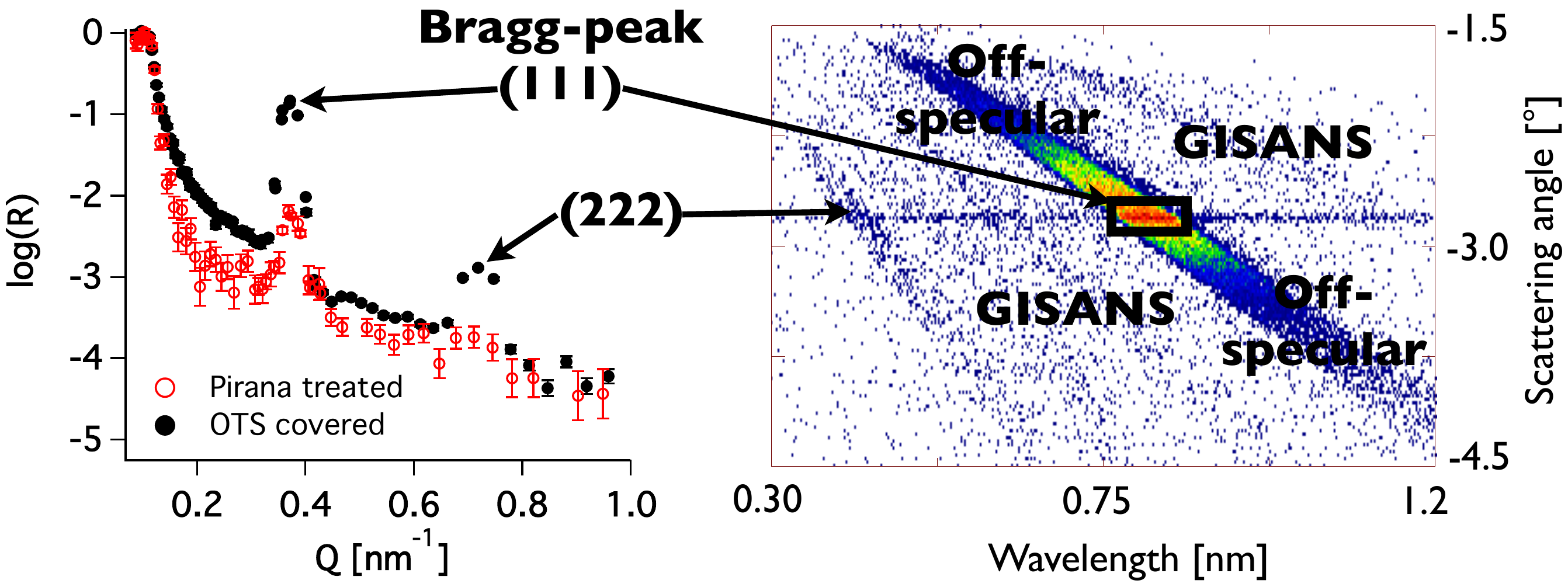}
\caption{\label{fig:refl}Neutron reflectivity of the sample measured in the crystalline phase (left panel). The right panel depicts raw data taken on FIGARO with the scattered intensity plotted versus wavelength and scattering angle.}%
\end{figure}

Figure \ref{fig:refl}, left panel, depicts the neutron reflectivity of a 20 \% solution of F127 in deuterated water taken on the reflectometer FIGARO at the ILL, Grenoble, France.

\begin{figure}
\includegraphics[width=300pt]{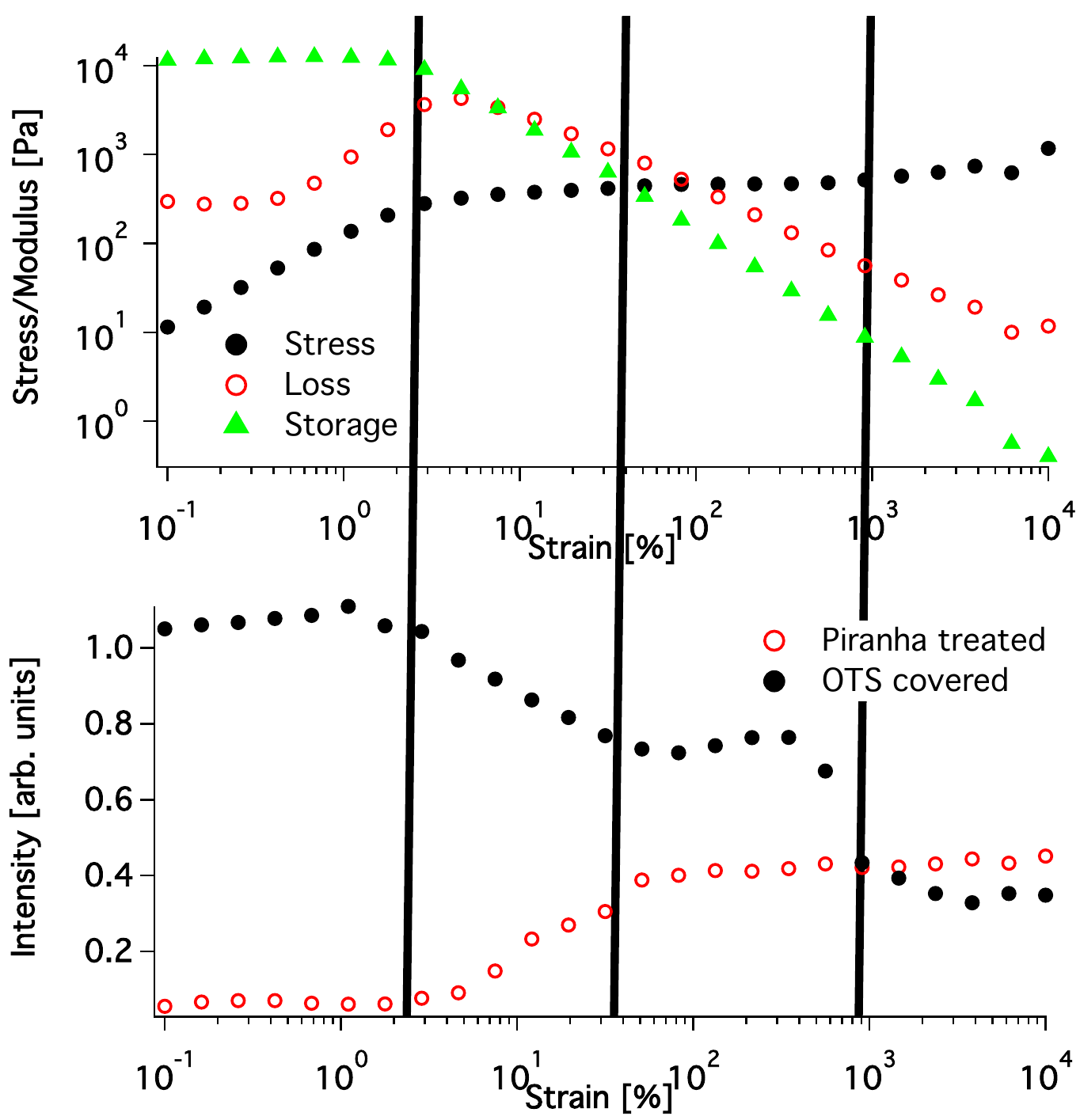}
\caption{\label{fig:osc}Stress, storage and loss modulus plotted together with the intensity of the Bragg reflection versus shear strain.}%
\end{figure}

The data are corrected for the the incident beam spectrum and the background is subtracted using the standard data processing software tool COSMOS \cite{campbell}, available on the instrument.
The reflectivity of both incident angles is shown after stitching them together.
In addition to the region of total external reflection visible at small Q values (below 0.1 nm$^{-1}$) two Bragg-reflections are found for data taken at 25 $^\circ C$ and with the sample in contact to the surface covered with OTS.
The first reflection is located at Q=0.37 nm$^{-1}$ and corresponds to a d spacing of 16 nm.
Assuming a cubic close packing of the micelles at the interface \cite{Wolff:2004p2390} the lattice parameter of the fcc structure is 29.5 nm and the two Bragg reflections can be identified as (111) and (222). 
A detailed discussion of the static arrangement of F127 micelles close to surfaces with different surface energies can be found in ref. \cite{Wolff:2004p2390}.
The right panel in figure \ref{fig:refl} depicts the scattered intensity plotted versus wavelength and scattering angle, as collected on FIGARO.
The intensity for an incident beam angle of 1.4$^\circ$ is shown without any correction.
The horizontal streak at 2.8$^\circ$ marks the specular reflected intensity, which is shown corrected and extracted in the left panel.
The high intensity at wavelength around 0.85 nm and a scattering angle of 2.8$^\circ$ refers to the specular component of the (111) Bragg reflection.
The strong off-specular scattering, diagonal streak, confirms a highly correlated roughness in the micellar crystal.
The weak diagonal streaks also visible in the intensity map are related to small angle scattering from the micellar crystal and are only visible since a relaxed resolution, divergence, is used on FIGARO parallel to the sample surface in order to gain in flux.
For a more detailed discussion of the off-specular and gracing incidence small angle scattering from static micellar crystals we refer the reader to ref. \cite{Wolff:2005p2521,Wolff:2009p64}.
In addition to the data taken with the sample in contact to the OTS coated silicon wafer the read circles represent the reflectivity in contact with the Piranha treated one.
A huge difference in the intensity of the (111) Bragg reflection is clearly visible.
Note, the measurement with the sample in contact to the Piranha treated surface was taken for a shorter time resulting in worse statistics and representing the quality of data achievable in the measurement time specified above.
In this article, we focus on the specular intensity of the (111) reflection for the sample exposed to an oscillatory shear force with a frequency of 1 Hz and as a function of deformation or strain.
The intensity, depicted in Figure \ref{fig:osc}, has been extracted from the detector pictures by integrating the scattered intensity in a certain solid angle and wavelength range, marked by the black box in figure \ref{fig:refl}, right panel.\\
Figure \ref{fig:osc} depicts the result in a graph where the storage and loss modulus are plotted together with the peak intensity, top and bottom panel, respectively.
Three regions can be distinguished.
At small strain values the storage modulus is larger than the loss modulus, implying a dominantly elastic behavior of the micellar crystal.
In this region we find a much lower intensity of the (111) reflection with the sample in contact with the Piranha treated silicon surface as compared to the oxide and OTS covered surfaces.
This fact indicates a surface induced ordering and is in line with previous studies \cite{Wolff:2004p2390}.
At strain values of about 1 \%, the loss modulus starts increasing and the storage modulus starts slowly decreasing.
At the same time, the intensity of the Bragg reflection starts to increase for the Piranha surface and to decrease for the other surface.
Above about 5 \% deformation, the loss modulus becomes larger than the storage modulus.
This indicates that the sample shows a more liquid-like behavior.
At the same time the intensities of the Bragg reflections remain unchanged, which can be explained by sliding of subsequent layers of micelles parallel to the solid surface.
Finally, at large deformations above 1000 \% there is a difference in storage and loss modulus of more than one order of magnitude and the intensities of the Bragg reflections at the different surfaces becomes almost the same.
This indicates that for large stress values the influence of the interface on the local ordering becomes less important as compared to the influence of the shear forces.
We note that the cone-plate distance is 0.1 mm in the center of the cone and the cone angle 1$^\circ$.
Comparing this to the penetration depth (depth at which the neutron beam is attenuated to 1/e) of the neutron beam which is on the order of some tens $\mu m$ it is clear that the rheometer in this configuration is not sensitive to the surface induced changes revealed by neutron reflectivity.

\section{Conclusion}

Our results demonstrate that close to the interface the anisotropy imposed by shear competes with that resulting from the presence of the solid boundary.
The transition from elastic to viscous shear can be observed in the changes of the crystal Bragg peak intensity.
The technique presented in this manuscript opens new possibilities for the study of the near surface properties of complex liquids.
By evaluating the full three dimensional data taken on time of flight instruments it is possible to retain the near surface structure in liquids spanning length scales from nm up to $\mu$m with in-situ deformation.

\begin{acknowledgments}

The authors acknowledge the help of R. Campbell and C. Halbert during the various neutron scattering experiments as well as the Swedish research council (travel grant no.: 90045401 and project grant: A0505501) and STINT (contract number: IG2011-2067) for financial support.
The authors thank the Partnership for Soft Condensed Matter at ILL.\\

\end{acknowledgments}

%\bibliography{RheoRefl}% Produces the bibliography via BibTeX.

\begin{thebibliography}{99}
\bibitem{comprod} Commercial materials, instruments and equipment are identified in this paper in order to specify the experimental procedure as completely as possible.
In no case does such identification imply a recommendation or endorsement nor does it imply that the materials, instruments, or equipment identified are necessarily best available for the purpose.
\bibitem{corrosive} Corrosive. Acid-resistant gloves, protective goggles, and lab coats must be worn handling the Piranha solution.
\end{thebibliography}

\end{document}